\documentclass[%
 reprint,
 amsmath,amssymb,
 aps,
floatfix,
]{revtex4-2}

\usepackage{xspace}

\usepackage{multirow}
\usepackage{graphicx}
\usepackage{dcolumn}
\usepackage{bm}
\usepackage{hyperref}
\hypersetup{
	colorlinks=true,        
	linkcolor=blue,         
	citecolor=blue,        
	urlcolor=blue,           
}
\usepackage{dsfont}
\usepackage{subfigure}
\usepackage{comment}
\usepackage{color}
\usepackage[normalem]{ulem}
\usepackage{soul}
\usepackage{float}
\usepackage{orcidlink}

\begin{document}

\title{Neutrino luminosity and the energy spectrum of a nova outburst
\footnote{Released on March, 1st, 2021}}

\author{Hao Wang\,\orcidlink{0009-0008-4586-8176}$^{1,2}$}
\author{Chunhua Zhu$^{1}$}%
 \email{Contact author: chuanhuazhu@sina.cn}
\author{Guoliang L\"{u}$^{1,2}$}%
 \email{Contact author: guolianglv@xao.ac.cn}
\author{Lin Li$^{1}$}
\author{Helei Liu$^{1}$}
\author{Sufen Guo$^{1}$}
\author{Xizhen Lu$^{1}$}

\affiliation{%
 $^{1}$School of Physical Science and Technology, Xinjiang University, Urumqi, 830046, China\\
 $^{2}$Xinjiang Observatory, The Chinese Academy of Sciences, Urumqi, 830011, China
}%


\date{January 31, 2025}

\begin{abstract}

The nova outburst can produce a large number of neutrinos, whether it is the nuclear reaction process during the explosion or the shock wave acceleration proton process. 
We study the low-energy nuclear and thermal neutrino luminosity of novae with CO white dwarf (WD) mass ranging from 0.6 to 1.1 $\rm M_{\odot}$ with different accretion rates $\dot{M}$, core temperatures $(T_{\mathrm{C}})$, and mixing degrees. 
We find that during the accretion phase, low-energy neutrinos are mainly produced by pp chains and plasma decay, and photon luminosity is greater than low-energy nuclear and thermal neutrino luminosity. During the thermonuclear runaway (TNR) phase, low-energy neutrinos are mainly produced by the CNO cycle and photon-neutrino interaction, and the low-energy nuclear and thermal neutrino luminosity far exceeds the photon luminosity.
We find that the more massive the WD, the shorter the cycle time and the higher the low-energy nuclear neutrino luminosity. The higher the accretion rate, the lower the low-energy nuclear neutrino luminosity. If the accretion mixing effect is not taken into account, the outburst interval becomes longer, and the low-energy nuclear neutrino luminosity will be increased. For the cooler nova model $(T_{\mathrm{C}}=1\times10^{7}\rm K)$, the low-energy nuclear neutrino luminosity will be lower during the accretion phase and higher at the TNR.  
We also predict the neutrino luminosity and energy spectrum of the upcoming recurrent nova T Coronae Borealis (T CrB). We estimate that the next T CrB outburst has a low-energy nuclear neutrino peak luminosity of $2.7\times10^{8}\ \rm L_{\nu,\odot}$ and a low-energy nuclear neutrino outburst duration of 88 days. 
In addition, we predict that the high-energy hadronic neutrino flux produced by T CrB nova cannot be observed by the current-generation IceCube.
\end{abstract}


\maketitle

\section{Introduction} \label{sec:intro}
Neutrinos are considered the most abundant particles in the Universe and play a crucial role in the evolution of stars\citep{Pandey2024}, for example, helium flashes \citep{Catelan1996}, thermal pulses \citep{Shi2020}, and cooling of white dwarfs (WDs) \citep{Fontaine2001} in low-mass stars, late stage evolution of high-mass stars \citep{Woosley2002}, supernova explosion \citep{Janka2017}, and the cooling of neutron stars \citep{Nomoto1981}. In addition, the study of neutrinos has significantly contributed to our comprehension of the fundamental properties of the Universe. It can probe neutrino interactions at high energies and explore deviations from the Standard Model of particle physics, for example, measuring the cross section of neutrino and nucleon interaction for energies larger than $\rm 1\ TeV$ \citep{Reno2023} and proving that leptoquarks can change the neutrino cross section \citep{Klein2019}. High-energy neutrinos are only produced at the same time as high-energy $\gamma$ rays through the hadron process, so detecting high-energy neutrinos can also determine the origin of high-energy $\gamma$ rays \citep{Song2023}. Meanwhile, the flavor composition of high-energy neutrinos can be a probe of neutrino magnetic moments \citep{Popov2024}.
Neutrinos carry invaluable information about the existence (or absence) of energetic protons and shed light on the location of the $\gamma$-ray production region \citep{Sahakyan2014}.

Neutrino detection is rapidly developing as a new astronomical messenger. 
Super-Kamiokande (SK) is a detector used to study neutrino Cherenkov radiation from different sources\citep{Fukuda2003}. It primarily detects and analyzes particle interactions within the energy range of several MeV to a few tens of TeV, such as solar neutrinos\citep{Abe2016} and supernova neutrinos\citep{Mori2022}.
The IceCube detector, the first of its kind with a gigaton-scale capacity, was primarily designed to observe neutrinos originating from the most violent astrophysical phenomena in our Universe \citep{IceCubeCollaboration2006}, such as blazars \citep{icecube2018, Tavecchio2018, Oikonomou2019, Cao2024}, $\gamma$-ray bursts \citep{Biehl2018}, black holes \citep{Ma2024, Velzen2024}, active galactic nuclei, \citep{Kurahashi2022} and tidal disruption events \citep{Winter2021}. These neutrinos usually have extremely high energies, ranging from TeV to PeV and beyond.

Novae are significant sites for the production of neutrinos \citep{Razzaque2010}. The novae are usually binary systems \citep{Warner1995}, and the primary star is a white dwarf. According to the type of companion star, it can be divided into two categories: one is the classical nova, where the companion star is the main sequence star, and the other is the symbiotic nova, where the companion star is the red giant (RG). When a sufficient number of layers of material have accumulated, it eventually triggers a thermonuclear runaway (TNR) explosion on the surface of the WD \citep{Hernanz2005} and significantly enhances the luminosity of the WD to approximately $\rm 10^{4-5}\ L_{\odot}$ \citep{Gomar1998, Hellier2001, Knigge2011}; such a violent nuclear reaction would produce a large number of electron neutrinos. Meanwhile, the shock wave from the nova outburst accelerates the surrounding particles to produce neutrinos. The shock wave can be categorized into two distinct types: internal and external shock waves. Internal shock waves can be attributed to the collision between slower-moving ejecta and faster-moving ejecta \citep{Metzger2015}. The fast-moving outflow has the potential to collide with a preexisting dense wind emanating from the RG star, resulting in the generation of an external shock \citep{Lu2011, Martin2013}. Symbiotics are therefore more likely to produce meaningful hadronic emission at their forward shocks. However, classical novae can and do produce hadronic emission at internal shocks \citep{Metzger2015, Ackermann2014, Cheung2016,  Franckowiak2018}. In the “hadronic” scenario, the particles accelerated by the shock wave undergo interactions with the surrounding matter [proton-proton (pp) or proton-$\gamma$ ($\rm p\gamma$) collisions], resulting in the generation of neutral and charged muons. These muons subsequently decay into high-energy $(\rm >1\ GeV)$ neutrinos and $\gamma$ rays \citep{Abbasi2022}. The maximum energies of high-energy particles will be contingent upon the duration of the nova, the efficiency of the acceleration mechanism, and the energy losses incurred during cooling \citep{Acciari2022}. The particles are accelerated by external shock waves to produce high-energy neutrinos, which typically occur in symbiotic novae \citep{DeSarkar2023}.

T Coronae Borealis (T CrB) is a classical symbiotic recurrent nova \citep{Ilkiewicz2023}. The primary star of this type of nova is in close proximity to the Chandrasekhar limit, and it must undergo accretion at exceedingly high rates, approximately $1.7 \times 10^{-8}\ \rm M_{\odot}\ yr^{-1}$, in order to rapidly accumulate critical-mass envelopes and consequently outburst with such frequency \citep{Kato1991}. In addition to T CrB, there are three with red giant donors in the ten known recurrent novae of the Galaxy: V745 Sco, RS Ophiuchus, and V3890 Sgr. The WD mass in T CrB is $1.2\pm0.2\ \rm M_{\odot}$, and the companion red giant has a mass of $1.12\pm0.23\ \rm M_{\odot}$. The average accretion rate $(\dot{M}_{\mathrm{a}})$ during the superactive state is $2\times10^{-8}\ \rm M_{\odot}\ yr^{-1}$ with a maximum of $4\times10^{-8}\ \rm M_{\odot}\ yr^{-1}$ \citep{Zamanov2023}. The distance from T CrB to Earth is approximately 914 pc. Many groups predicted it will outburst in 2024 \citep{Schaefer2023, Maslennikova2023}. Prior to this, the T CrB is recorded eruptions in 1217.8, 1789.9, 1866.4, and 1946.1; the time span between bursts is approximately 80 yr \citep{Schaefer2023}. Because T CrB is very close to us and is about to outburst, it is necessary to study its neutrino luminosity and energy spectrum.

In this paper, we plan to investigate neutrino luminosity and energy spectrum during nova outburst. 
Sec. \ref{sec:2} gives the details of software instruments, input physics, reaction networks,
and model selection. In Sec. \ref{sec:3} we discuss the low-energy nuclear and thermal neutrino luminosity during nova outburst and the influence of different $M_{\rm{WD}}$, 
$\dot{M}$, and $T_{\mathrm{C}}$ on low-energy nuclear neutrino luminosity, we also predict the neutrino energy spectrum 
when the T CrB next outbursts. The conclusions are given in Sec. \ref{sec:4}.

\section{Nova MODEL} \label{sec:2}
Classical novae eruptions are the result of unstable hydrogen burning on the surface of either CO WD or ONeMg WD cores,
which accrete hydrogen-rich materials from their main sequence or RG companions in low-mass, close binary systems.
We employ the open-source stellar evolution code
\textsc{mesa}, version 23.05 (Paxton $et\ al$. \citep{paxton2011, paxton2013, paxton2015, paxton2018, paxton2019} and \citet{Jermyn2023}) to construct nova models.
\textsc{mesa} is capable of generating state-of-the-art simulations of novae \citep{Denissenkov2013, Zhu2021}.

In the nova model, the occurrence of mass loss is triggered when the luminosity ($L$) of a star surpasses the super-Eddington luminosity $(L_{\mathrm{Edd}})$. The mass loss rate is 
\begin{align}
     \dot{M}=-2 \eta_{\mathrm{Edd}} \frac{\left(L-L_{\mathrm{Edd}}\right)}{v_{\mathrm{esc}}^2} ,
\end{align}
where $L_{\mathrm{Edd}}=(4\pi \mathrm{Gc}M)/\kappa$, $v_{\mathrm{esc}}=\sqrt{2\mathrm{G}M/R}$. $M$ and $R$ are the mass and radius of the WD, while $\kappa$ is the Rosseland mean opacity at the WD’s surface. The scaling factor is taken as $\eta_{\mathrm{Edd}}=1$ \citep{Denissenkov2013}. The TNR begins when the total white dwarf luminosity $(L)$ exceeds $10^{4}$ times the solar luminosity $\rm (L_{\odot})$ and ends when it falls below $\rm 10^{3}L_{\odot}$ \citep{Lu2020, gao2024}. The opacities are from \citet{Iglesias1996}.
The three factors that determine the physical characteristics of nova eruption are the mass of the white dwarf $(M_{\mathrm{WD}})$, accretion rate $\dot{M}$, and core temperature $(T_{\mathrm{C}})$ of the WD \citep{Shara1980, Prialnik1995}.
The $M_{\mathrm{WD}}$ affects peak energy generation; the temperature in the nuclear burning region increases as the $M_{\mathrm{WD}}$ increases \citep{Starrfield2020}.
The $\dot{M}$ will influence the efficiency of compression heating, thereby impacting the time to thermonuclear runaway, the mass of accretion, and the intensity of the outburst \citep{Prialnik1982}.
The influence of $T_{\mathrm{C}}$ on nova outburst can be divided into two cases \citep{Schwartzman1994}: In cold WD $(T_{\mathrm{C}}\leq1\times10^{7}\rm K)$, the accretion layer of the WD will also be cold, resulting in increased outburst interval and more mass of accretion material. In hot WD $(T_{\mathrm{C}}\geq3\times10^{7}\rm K)$, convection generated by high temperatures in the outer core region can accelerate the mixing of accretion material with core material. 
Relevantly, the consideration of these three parameters becomes imperative in talking about neutrino luminosity.
So our models set the $T_{\mathrm{C}}=1\times10^{7}, 2\times10^{7}$, and $3\times10^{7}\rm\ K$ and select the CO model mass as $0.6\ \rm M_{\odot}$, $0.7\ \rm M_{\odot}$, $0.8\ \rm M_{\odot}$, $0.9\ \rm M_{\odot}$, $1.0\ \rm M_{\odot}$, and $1.1\ \rm M_{\odot}$ and the ONeMg model mass as $1.1\ \rm M_{\odot}$, $1.3\ \rm M_{\odot}$. 
We set the accretion rates at $1\times10^{-8} \sim\ 1\times10^{-10}\ M_{\odot}\rm \ yr^{-1}$, which are compatible with observations \citep{Patterson1984}.
In addition to the three parameters mentioned above, the solarlike material transferred from the companion will undergo mixing with the outer layers of the WD before the nova outburst \citep{Starrfield2016, Rukeya2017}.
In \textsc{mesa}, we replace the mixing effect by changing the composition of the accreting material; that is, 50\% of WD material and 50\% of solar material, or 25\% of WD material and 75\% of solar material \citep{Lodders2009}. According to \citet{Denissenkov2014}, we take the latter as our mixing nova model.

The neutrino normalization $L_{\nu, \odot}=0.02398 \cdot L_{\gamma, \odot}=9.1795 \times 10^{31}\ \mathrm{erg\ s}^{-1}$ is adopted in \citet{Farag2020}.

\subsection{Neutrinos produced by nuclear reaction and thermal processes}
\label{subsection:2.1}
The generation of neutrinos inside stars can be broadly categorized into two distinct processes.
The first includes weak nuclear processes, including $\beta^{\pm}$ decay, electron capture (EC), and positron capture (PC) \citep{Kato2020},

\begin{align}
	&&& \rm nuclei\ EC\ :\ (Z, A)+e^{-} \longrightarrow(Z-1, A)+\nu_{e} \\
	&&& \rm free\ proton\ EC\ :\ p+e^{-} \longrightarrow n+\nu_{e} \\
	&&& \rm PC\ :\ (Z, A)+e^{+} \longrightarrow(Z+1, A)+\bar{\nu}_{e} \\
	&&& \rm \beta^+\ :\ (Z, A) \longrightarrow(Z-1, A)+e^{+}+\nu_{e} \\
	&&& \rm \beta^-\ :\ (Z, A) \longrightarrow(Z+1, A)+e^{-}+\bar{\nu}_{e}
\end{align}

such as $\rm p(p,e^{+}\nu)d$, $\rm p(e^{-}p,\nu_{e})d$, and $\rm ^{3}He(p,e^{+}\nu_{e})\alpha$ in pp chains and $\rm ^{13}N(,e^+\nu_{e})^{13}C$, $\rm ^{15}O(,e^+\nu_{e})^{15}N$, and $\rm ^{17}F(,e^+\nu_{e})^{17}O$ in the CNO cycle. For the nova model, we chose pp-and-cno-extras.net as a network of nuclear reactions during nova outburst. The occurrence of classical novae eruptions can be attributed to the instability in hydrogen burning on the surface of WD cores. Therefore, the most important nuclear reactions during nova eruption are the pp chain and the CNO cycle \citep{Denissenkov2013}. 

The other includes thermal processes, which is sensitive to temperature and density. Thermal neutrino energy losses are from \citet{Itoh1996}, including the following:
\vspace{0.2cm} 

\noindent pair annihilation$\rm \left(T>10^9\ K\right)$ 
\begin{align}
	 e^{+}+e^{-} \longrightarrow \nu+\bar{\nu}
\end{align}
photon-neutrino interaction$\rm \left(T<4\times10^8\ K, \rho<10^5\ g\ cm^{-3}\right)$ 
\begin{align}
	 e^{-}+\gamma \longrightarrow e^{-}+\nu+\bar{\nu}
\end{align}
plasma decay$\rm \left(10^7<T<10^8\ K, 10^4<\rho<10^7\ g\ cm^{-3}\right)$ 
\begin{align}
	 \rm \gamma^{*} \longrightarrow \nu+\bar{\nu}
\end{align}
and bremsstrahlung$\rm \left(10^8<\rho<10^{10}\ g\ cm^{-3}\right)$
\begin{align}
	 e^{-}+(Z, A) \longrightarrow e^{-}+(Z, A)+\nu+\bar{\nu}
\end{align} 

The \textsc{mesa} nuclear reaction rates are a combination of rates from NACRE \citep{Angulo1999} and JINA REACLIB \citep{Cyburt2010} databases. 
The treatment of screening corrections, as proposed by \citet{Chugunov2007}, serves to enhance nuclear reaction rates in dense plasmas. 
The weak reaction rates utilized in this study are derived from previous works \citep{Langanke2000, Oda1994, Fuller1985}.

\subsection{Neutrinos produced by $pp$ collisions}
T CrB is a recurrent nova; the ambient environments and explosion mechanisms are similar to RS Ophiuchus (RS Oph) \citep{Evans2019}. When the ejection material encounters the wind material around the red giant, it will cause a shock system \citep{Zamanov2023}.

The interaction between RG wind and the ejecta will produce two shocks, a forward shock propagating into the RG wind and a reverse shock propagating back into the ejecta \citep{Zheng2022}.
According to \citet{Aharonian2022}, we also only consider the forward shock for explaining the neutrino emission.

In the early stage, the ejecta expands freely and is almost unaffected by the interstellar medium surrounding the binary. We assume 
when the distance between the wind and the nova $r$ is much larger than the semimajor axis of the nova $a$, the structure of the wind can be approximately spherical, and the ejecta is assumed to be concentrated at the front of the shock wave with a thickness of $h r_{\mathrm{sh}}$,
where $h$ is $1/10$ of the shock radius.
Meanwhile, we are only interested in the initial day of the T CrB outburst, thus we do not consider the radiative cooling phase after the shock wave deceleration. We assume that the shock radius $(r_{\mathrm{sh}})$ and velocity $(v_{\mathrm{sh}})$ of the T CrB are $\rm 4\times10^{13}\ cm$ and $\rm 4500\ km/s$ \citep{Zheng2024}, respectively. The T CrB companion has a mass of $1\ \rm M_{\odot}$ \citep{Pavlenko2020} and a typical mass loss rate $(\dot{M}_{\mathrm{R G}})$ of $5\times10^{-7}\ \rm M_{\odot}\ \mathrm{yr}^{-1}$ \citep{Ferrarotti2006}.
The velocity of RG wind $(v_{\mathrm{RG}})$ is $\rm 10\ km/s$.

The accelerated protons in the forward shock will interact with the matter in the RG wind, mainly the collision of protons and protons, and the resulting charged pion and kaon decay produces observable $\gamma$-ray and high-energy neutrinos. The density of the matter in the RG wind can be estimated as

\begin{align}
    &&& n_{\mathrm{R G}}  =\frac{\dot{M}_{\mathrm{R G}}}{4 \pi r_{\mathrm{s h}}^2 v_{\mathrm{R G}} m_{\mathrm{p}}} =9.9 \times 10^8 \mathrm{~cm}^{-3}\ ,
\end{align}
where $m_{\mathrm{p}}$ is the rest mass of the proton.
 
The neutrino energy spectrum is also sensitive to the maximum accelerated proton energy, and it is to relate the acceleration timescale with the age of the system. If particles are accelerated via diffusive shock acceleration, then the acceleration time should be of order of the diffusion time ($D/v_{\mathrm{sh}}^2$). For Bohm diffusion, the characteristic timescale of particle acceleration can be estimated as

\begin{equation}
    \tau_{\text {acc }}= c R_{\mathrm{L}}/3 \xi v^{2}_{\mathrm{sh}} \ \mathrm{~s}\ ,
\end{equation}
where $\xi \leq (v_{sh}/c)^{2} \approx 2 \times 10^{-4}$ is acceleration parameter, $R_{\mathrm{L}}=E/eB$ is the Larmor radius of particles in the magnetic field B (in Gauss). According to \citet{Bednarek2011} and \citet{Marcowith2018}, the proton Larmor radius for a particle energy of 1 GeV can be written as 

\begin{equation}
    R_{\mathrm{L}} \simeq 3 \times 10^{6} E_{\mathrm{GeV}} B_{\mathrm{G}}^{-1} \mathrm{~cm}\ .
\end{equation}

As discussed in \citet{Chomiuk2012}, the magnetic field B is

\begin{align}
    B=\sqrt{32 \pi n_{\mathrm{R G}} k_{\mathrm{B}} T_{\mathrm{R G}}}\ ,
\end{align}
where $T_{\mathrm{RG}}\rm$ is the temperature of the RG wind, $k_{B}=1.380649 \times 10^{-23}\ \rm J/K$. Based on \citet{Stanishev2004}, the RG companion star of T CrB has an effective surface temperature of 3500 K. However, the $T_{\mathrm{RG}}\rm$ produced
from the companion star should be less than the surface temperature \citep{Chomiuk2012}. Following \citet{DeSarkar2023} for nova RS Oph, we set $T_{\mathrm{RG}}=1000\rm\ K$.
The primary mechanism for hadron energy loss occurs through collisions with the matter present in RG winds. An estimation of the timescale for energy losses due to pion production in $pp$ collisions can be derived from \citep{Diesing2023},

\begin{equation}
    \tau_{p p}=\left(\sigma_{p p} k \mathrm{c} n_{\mathrm{R G}}\right)^{-1} \approx 2.2 \times 10^7 (\frac{10^8\ \mathrm{cm}^{-3}}{n_{\mathrm{R G}}})\ \mathrm{s}\ ,
    \end{equation}
where $\sigma_{p p}\approx3\times10^{-26}\rm\ cm^{2}$ is the total cross section of $pp$ interactions, and $k = 0.5$ is the inelasticity coefficient in this collision. By comparing the timescale for the energy losses in $pp$ collisions and the characteristic acceleration timescale ($\tau_{\mathrm{pp}}=\tau_{\mathrm{acc}}$), the maximum proton energy can be estimated as 

\begin{equation}\label{equ:8}
    E_{p, \max } \approx 1 \times 10^{-6}\frac{\xi B \tau_{\mathrm{pp}} v_{\mathrm{sh}}^2}{c} \ \rm GeV\ .
    \end{equation}

According to \citet{Caprioli2023}, the diffusion length ($D/v_{\mathrm{sh}}$) of the particles with energy $E_{max}$ cannot exceed the source size $r_{\mathrm{sh}}$, which is similar to the Hillas criterion. For our hadronic model, we find that the accelerated proton population can reach a maximum energy of approximately 330 GeV. As described by \citet{Caprioli2010}, the maximum energy is usually the location of the cutoff energies, so we set $E_{0}=330\rm\ GeV$. The results obtained from the nova model discussed in this section will be presented in the subsequent section.

\section{Result} \label{sec:3}
Using \textsc{mesa}, we simulate the structure and evolution of
seven nova models including different initial CO WD masses from 0.6 to 1.1 M$_\odot$ with a mass interval of 0.1 M$_\odot$, and 1.1 M$_\odot$ ONeMg WD. 
Following \citet{Farag2020}, we utilize the neutrino H-R diagram to illustrate the neutrino luminosities throughout the complete life cycles of nova outburst
and discuss how $M_{\mathrm{WD}}$, $\dot{M}$, $T_{\mathrm{C}}$, and mixing affect low-energy nuclear neutrino luminosity of the nova.

\subsection{Neutrino luminosity of nova outburst}
\begin{figure*}
    \centering
        \includegraphics[width=0.8\textwidth,height=0.8\textwidth]{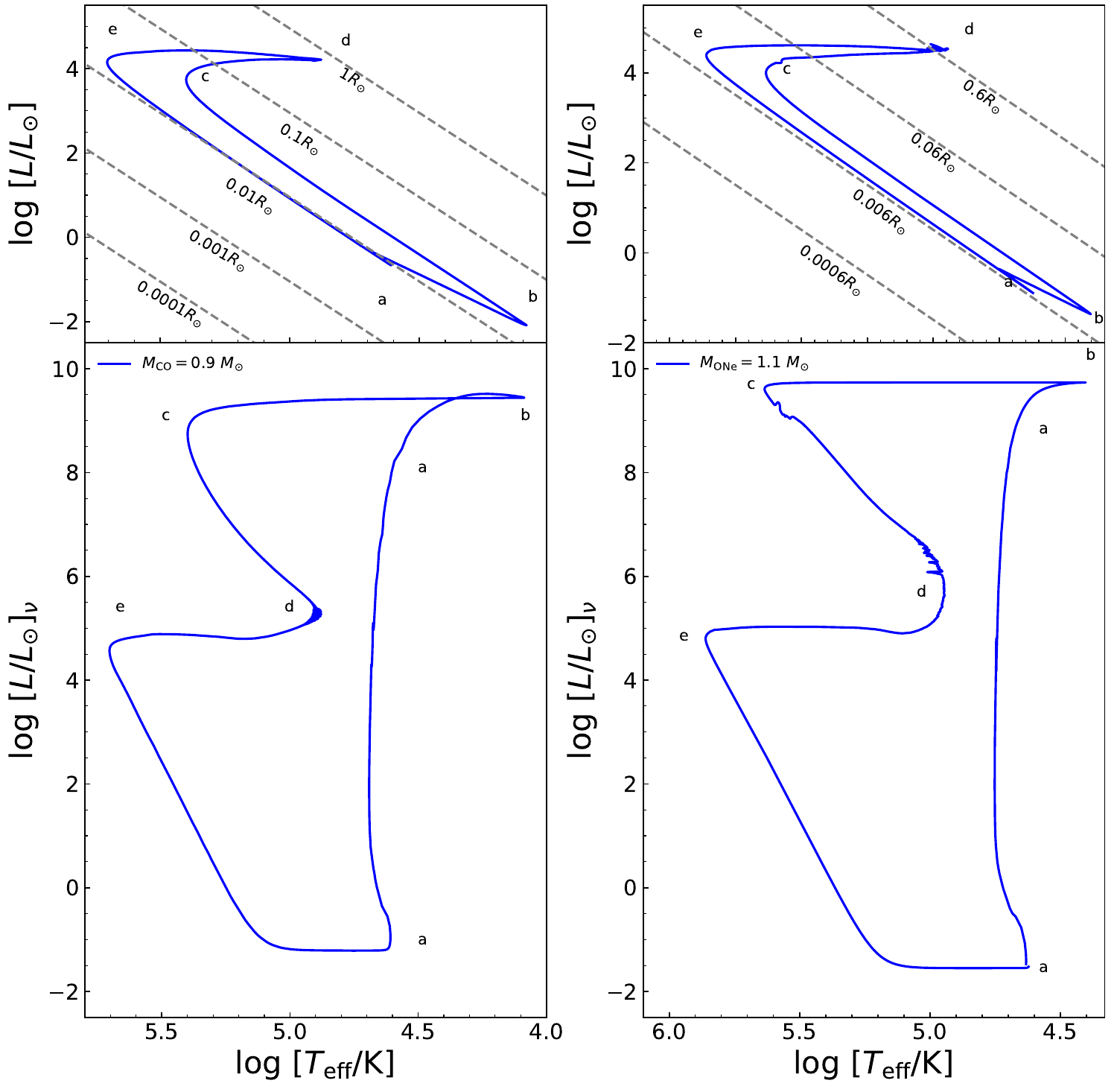}    
        \caption{The track in the H-R diagram and neutrino H-R diagram for the whole nova multicycle evolution. The left is 0.9 M$_\odot$ CO nova model, the right is 1.1 M$_\odot$ ONe nova model. Both have center temperature $T_{\mathrm{C}}=3\times10^{7}\ \rm K$ and accrete $25\%$ WD material and $75\%$ solar material at a rate $1\times10^{-9} \ \mathrm{M}_{\odot}\rm\ yr^{-1}$.}
        \label{fig:1}
    \end{figure*}

Figure \ref{fig:1} shows the nova with 0.9 M$_\odot$ CO WD and 1.1 M$_\odot$ ONeMg WD singly periodic evolutionary tracks in both the H-R diagram and the neutrino H-R diagram.  
Several key moments during the nova eruption have been marked in the H-R diagram and neutrino H-R diagram. The $a\to b\to c\to d$ is the explosion curve, and the $d\to e\to a$ is the cooling curve. The $a\to c$ clearly illustrates that, when the TNR occurs, due to the rapid rise of temperature, the intense nuclear reaction produces a large number of neutrinos, the cross section of neutrinos interacting with the matter is extremely low compared to photons, and the low-energy nuclear and thermal neutrino luminosity rises rapidly in a straight line, while photon luminosity remains almost constant. This is why, in the neutrino H-R diagram, there are two points labeled $a$. When the low-energy nuclear and thermal neutrino luminosity reaches its maximum, it also means that the nuclear reaction rate on the surface of the WD reaches its maximum, the photon luminosity begins to rise, and low-energy nuclear and thermal neutrino luminosity would remain constant. 
The $c\to d$ indicates that when part of the photon's energy is injected into the shell, the shell begins to expand, and the surface temperature of the WD decreases. The particle number density in the nuclear reaction zone decreases, and the nuclear reaction rate begins to decrease, so the low-energy nuclear and thermal neutrino luminosity begins to decrease.
The $d\to e$ indicates that photosphere of the WD begins to contract, the effective temperature rises, and the nuclear reaction rate remains constant, as does the photon luminosity and low-energy nuclear and thermal neutrino luminosity.
Finally, the nova outburst is over. The cooling process of the WD commences as the nuclear reaction concludes, leading to a gradual decline in photon and low-energy nuclear and thermal neutrino luminosity. 

\begin{figure*}
    \centering
        \includegraphics[width=\textwidth,height=0.4\textwidth]{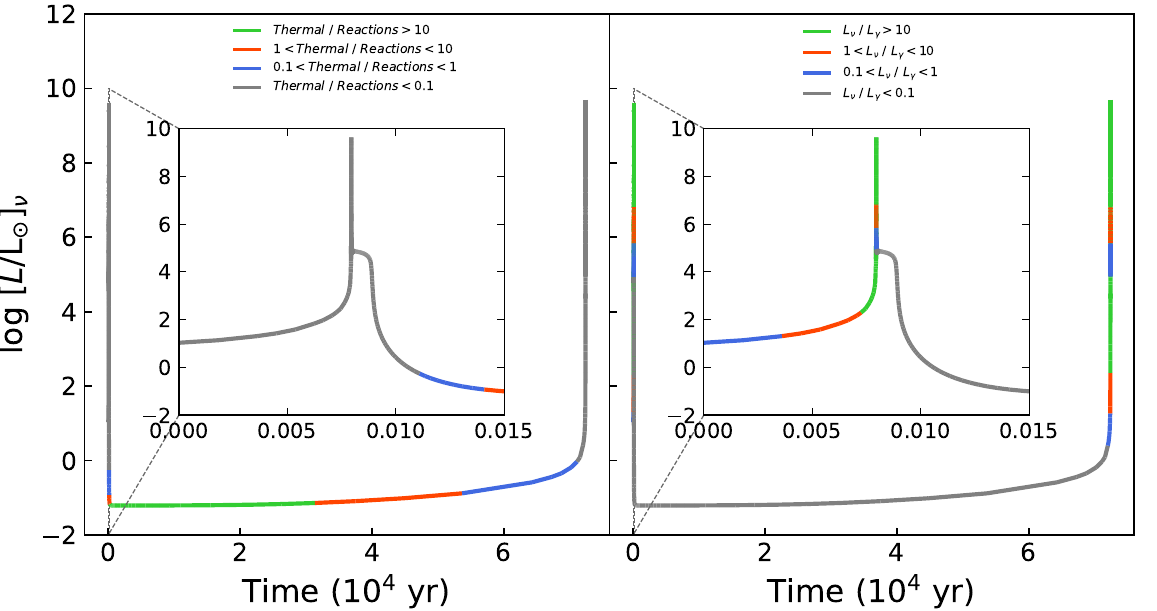}    
        \caption{The low-energy neutrino luminosity produced by the nuclear reaction and thermal processes (left).
		The colorful solid lines are the ratio of the low-energy nuclear reaction neutrino luminosity to the low-energy thermal
		neutrino luminosity plotted along the light curve of the nova eruption.
		Gray curves indicate where nuclear reaction neutrinos dominate,
		green curves where thermal neutrinos dominate, and blue and red curves where the
		reaction and thermal neutrino luminosities are within a factor of 10.
		Right: the ratio of the low-energy nuclear and thermal neutrino luminosity to the photon luminosity during the whole nova multicycle evolution.
		The colorful solid lines give the ratio
		of the low-energy nuclear and thermal neutrino luminosity to the photon luminosity in a light curve of the nova eruption. The nova model has a 0.9 $\rm M_{\odot}$ CO WD,
		center temperature $T_{\mathrm{C}}=3\times10^{7}\ \rm K$, and accretes $25\%$ WD material and $75\%$ solar material at a rate $1\times10^{-9} \ \mathrm{M}_{\odot}\rm\ yr^{-1}$.}
        \label{fig:2}
    \end{figure*}

The production of neutrinos can be achieved through both nuclear reactions and thermal processes, as discussed in Sec. \ref{subsection:2.1}.
The left panel of Fig. \ref{fig:2} illustrates the ratio of the low-energy nuclear reaction neutrino luminosity to the low-energy thermal process neutrino luminosity, traced along the light curve of the nova eruption.
The right panel of Fig. \ref{fig:2} shows the ratio of the low-energy nuclear and thermal neutrino luminosity to the photon luminosity, traced along the light curve of the nova eruption.
During the accretion phase, the temperature of the accretion layer is less than the Fermi temperature $(7\times10^{7}\rm\ K)$, the nuclear reactions are weak, neutrinos are mainly produced by thermal processes, and the accretion layer of the WD radiates energy mainly through photons.
However, when the TNR occurs, the neutrinos produced by nuclear reactions will far exceed those produced by thermal processes. Thus, neutrinos loss becomes the main channel of energy loss at the TNR. In addition, low-energy nuclear and thermal neutrino luminosity far exceeds photon luminosity when the nova outbursts. For example, as our $M_{\mathrm{CO}}=0.9\ \rm M_{\odot}$ nova model, when it outbursts, the peak photon luminosity is $5.5\times10^{4}\ \rm L_{\odot}$, and the peak low-energy nuclear neutrino luminosity is up to $3.6\times10^{6}\ \rm L_{\gamma,\odot}$.

Since the low-energy nuclear neutrino luminosity is much larger than low-energy thermal neutrino luminosity during nova outbursts, we only discuss the low-energy nuclear neutrino luminosity during nova outburst later. 

\begin{figure*}
    \centering
        \includegraphics[width=\textwidth,height=0.4\textwidth]{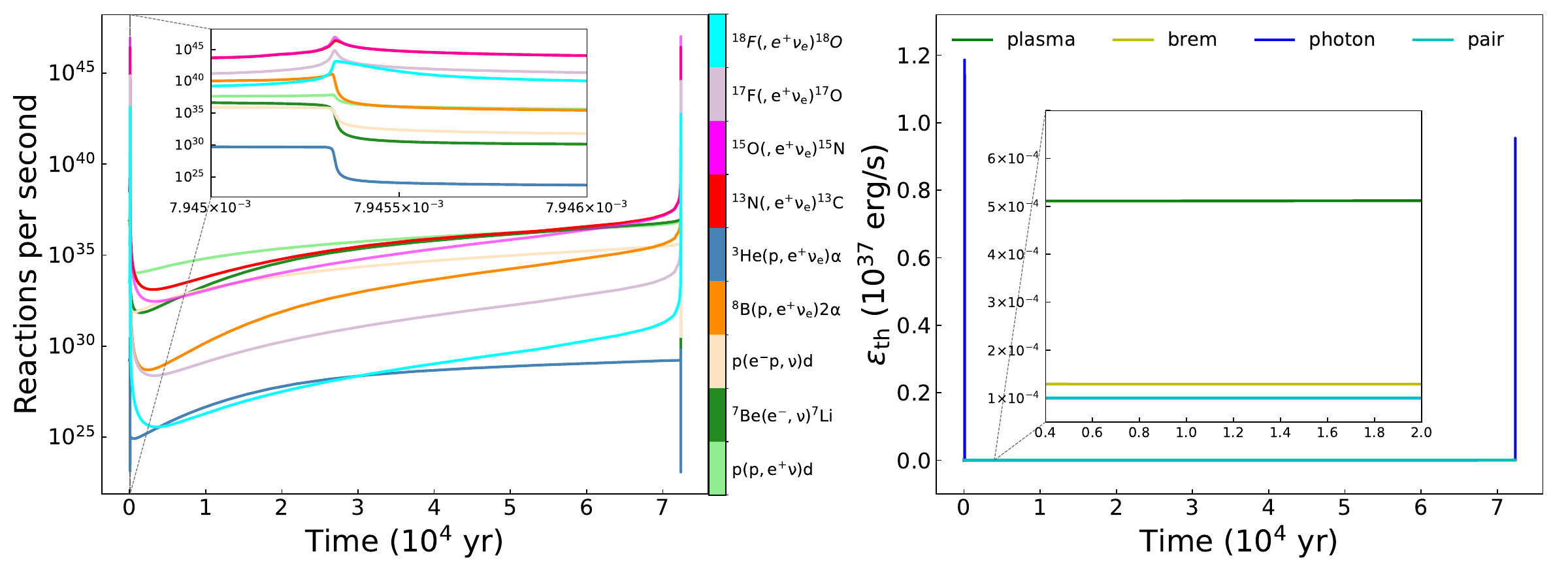}    
        \caption{The nuclear reactions rates of the nine neutrinos producing nuclear reactions in the pp chain and the CNO cycle during the whole nova multicycle evolution (left).
		Right: the energy yield of four thermal processes during the whole nova multicycle evolution.
		The blue line is plasma decay, yellow line is bremsstrahlung, green line is the photon-neutrino interaction, cyan line is electro-pair annihilation. 
		The nova model is similar to Fig. \ref{fig:2}.}
        \label{fig:3}
    \end{figure*}

Figure \ref{fig:3} shows the energy yields of the four thermal neutrino processes and the nuclear reaction rates of the nine neutrinos producing nuclear reactions in the pp chain and the CNO cycle.
The left panel shows that when the pre-TNR accreted mass is determined, for the nuclear process, the proton-proton chain plays a crucial role during the primary accretion phase, especially the $pp$ reaction: $\rm p+p \to d+e^{+}+\nu$ \citep{Starrfield2016} and $\rm ^{7}Be$ electron capture :$\rm ^{7}Be(e^{-},\nu)^{7}Li$. As accretion progresses, when the temperature of layers is about $7\times10^{7}\ \rm K$, the degeneracy is no longer important and the layers begin to expand, the temperature begins to rise rapidly, nuclear reaction neutrinos are dominated by the CNO cycle, and the TNR occurs. Neutrinos are produced mainly by four nuclear reactions: $\rm ^{13}N(e^+,\nu_{e})^{13}C$, $\rm ^{15}O(e^+,\nu_{e})^{15}N$, $\rm ^{17}F(e^+,\nu_{e})^{17}O$, and $\rm ^{18}F(e^+,\nu_{e})^{18}O$. 
For the thermal process, during the accretion phase, the WD has a high density in both its interior and exterior layers, the neutrino production is dominated by plasma decay. Once the TNR occurs, neutrino production will be dominated by photon-neutrino interaction because the temperature of layers is in the temperature range $\rm (T<4\times10^8\ K)$ that it dominates, as shown in the right panel. 

\subsection{Neutrino luminosity of novae with different $M_{\mathrm{WD}}$, accretion rate $\dot{M}$, $T_{\mathrm{C}}$, and mixing}
\begin{figure}
    \centering
        \includegraphics[width=\columnwidth]{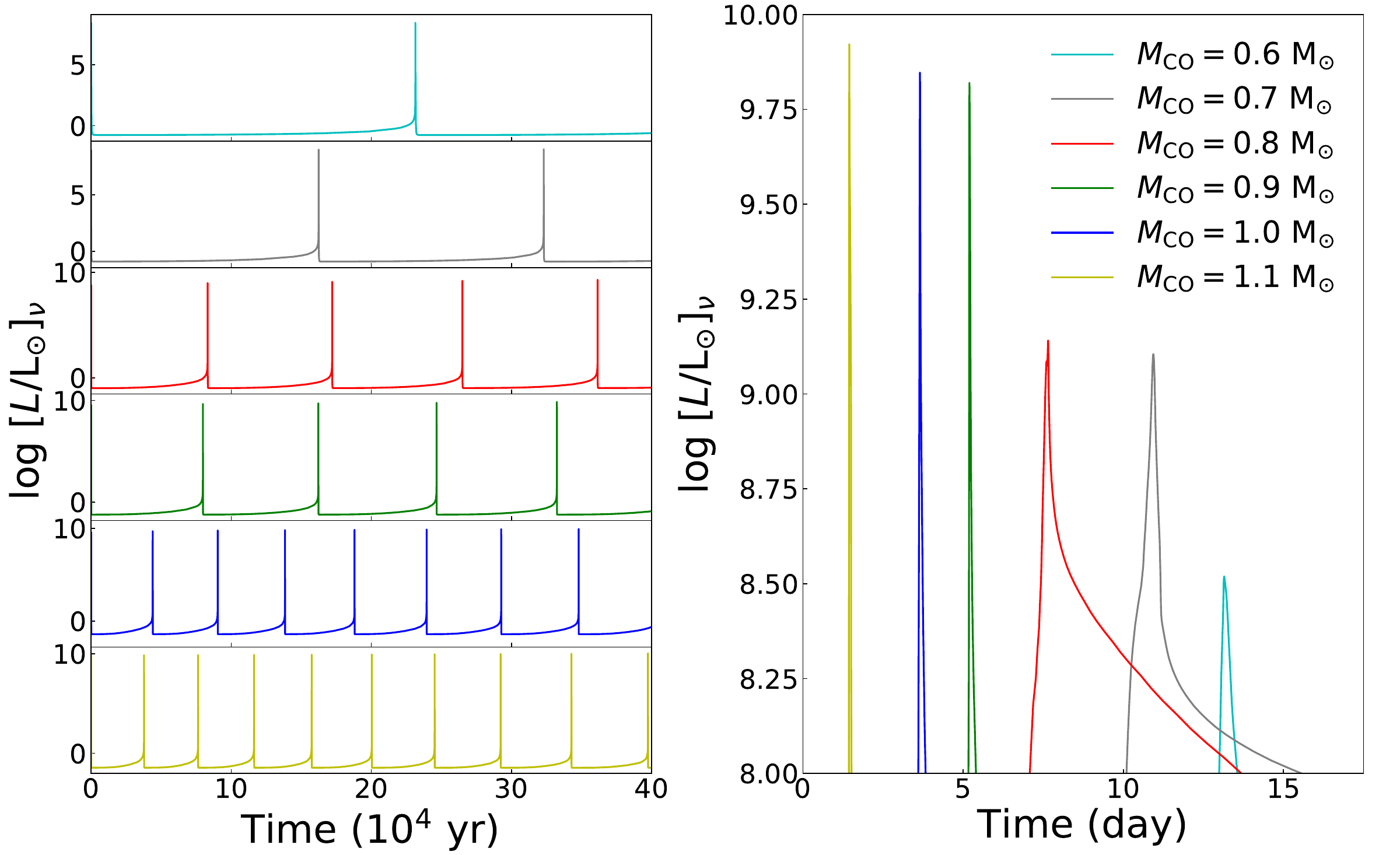}    
        \caption{Neutrino luminosity curves with different CO nova mass models. Left:
		the neutrino luminosity of the burst sequence during 0$-$400000 yr with different CO nova mass models, i.e., 0.6 $\rm M_{\odot}$ (cyan), 0.7 $\rm M_{\odot}$ (gray), 0.8 $\rm M_{\odot}$ (red), 0.9 $\rm M_{\odot}$ (green), 1.0 $\rm M_{\odot}$ (blue), and 1.1 $\rm M_{\odot}$  (yellow).  Right: average light curves.
		Both nova models have center temperature $T_{\mathrm{C}}=3\times10^{7}\ \rm K$, and accrete $25\%$ WD material and $75\%$ solar material at a rate $\dot{M}=1\times10^{-9} \ \mathrm{M}_{\odot}\rm\ yr^{-1}$.}
        \label{fig:4}
    \end{figure}

The $M_{\mathrm{WD}}$ of the nova can affect temperatures in the nuclear burning region.
The left panel of Fig. \ref{fig:4} shows the neutrino luminosity curves of the range of CO $M_{\mathrm{WD}}$ is $0.6\ \rm M_{\odot} - 1.1\ \rm M_{\odot}$.  
The right panel shows the average peak low-energy nuclear neutrino luminosity of ten outbursts.
It is clearly illustrates that the peak luminosity of neutrinos increases as the mass of the WD increases. 
The larger the mass and smaller the radius of a WD, the greater the gravitational acceleration. This makes it harder for hydrogen to penetrate the degenerate WD core. That means a more degenerate hydrogen-rich envelope, higher temperatures, and, of course, a more violent explosion.
As discussed above, the nuclear neutrino is mainly produced by the CNO cycle, which is highly sensitive to template ($\varepsilon_{\rm CNO} \propto T^{17}$).
So the more massive the $M_{\mathrm{WD}}$, the higher the low-energy nuclear neutrino luminosity during the TNR.

\begin{figure}
	\centering
		\includegraphics[width=\columnwidth]{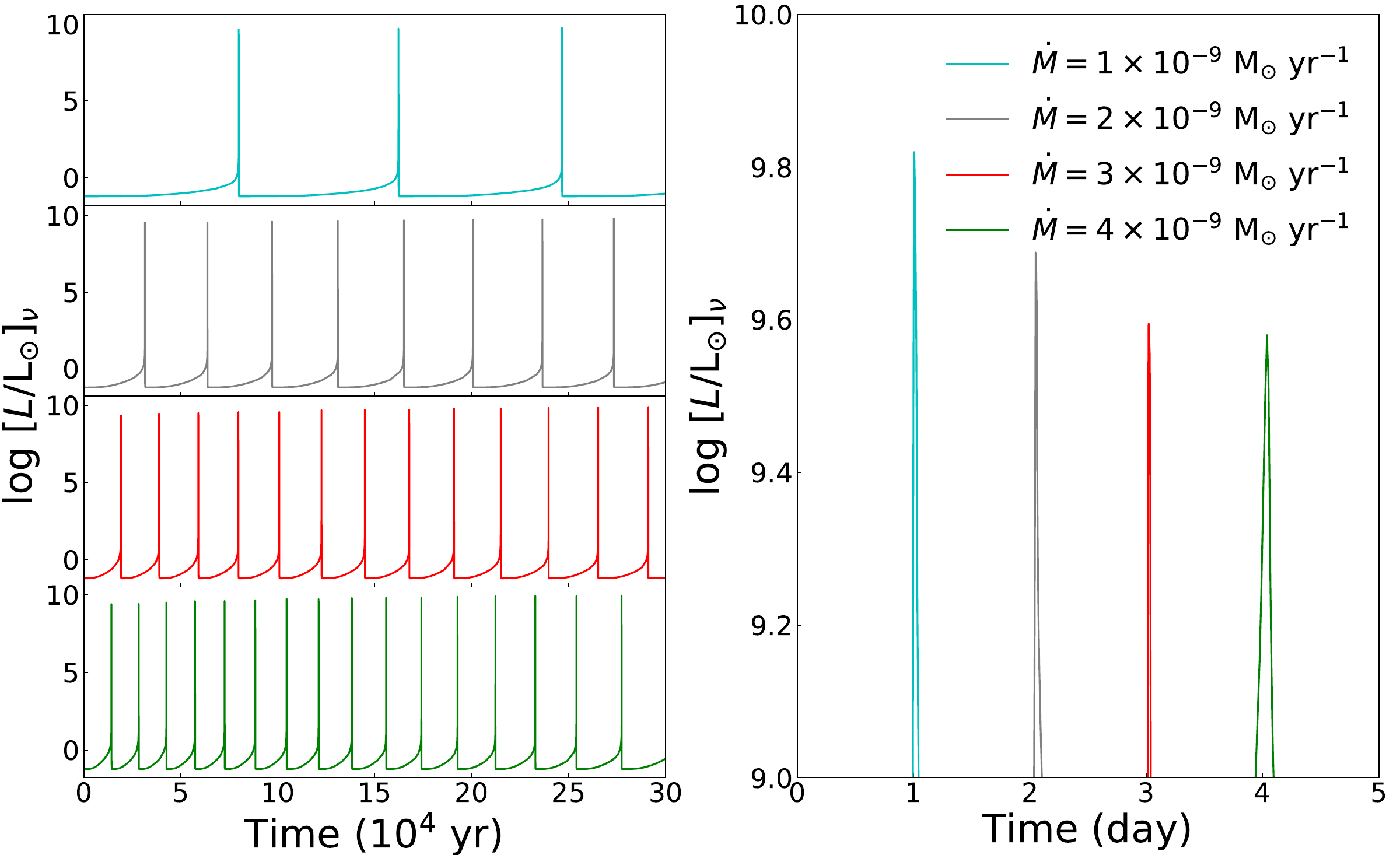}    
		\caption{Neutrino luminosity curves with different $\dot{M}$. Left:
		the neutrino luminosity of the burst sequence during 0$-$300000 yr with different $\dot{M}$, i.e., $1\times10^{-9}$ (cyan), $2\times10^{-9}$ (gray), $3\times10^{-9}$ (red), $4\times10^{-9} \ \mathrm{M}_{\odot}\rm\ yr^{-1}$ (green). Right: average light curves.
		Both nova models have center temperature $T_{\mathrm{C}}=3\times10^{7}\ \rm K$, and accrete $25\%$ WD material and $75\%$ solar material at a 0.9 $\rm M_{\odot}$ CO WD.}
		\label{fig:5}
	\end{figure}

The left panel of Fig. \ref{fig:5} shows the neutrino luminosity curves of nova models with different $\dot{M}$. The right panel shows the average peak low-energy nuclear neutrino luminosity of ten outbursts.
It clearly shows that the higher the accretion rate $\dot{M}$, the smaller the outburst interval $\Delta t_{\mathrm{rec}}$ and $L_{\nu}$.
TNR is produced when the hydrogen-rich material is heated to a high enough temperature $\rm (7\times10^{7}\ K)$ for the hydrogen to burn through the CNO cycle. The high accretion rate provides heat at a high rate \citep{Prialnik1982}, thereby inducing an acceleration in nova eruptions. However, owing to the reduced mass of accretion, the lower TNR peak temperature eventually gives rise to a diminished low-energy nuclear neutrino luminosity at higher accretion rates compared to lower ones.
When the accretion rate is low, the accretion layer temperature rises slowly and the outburst interval is larger, which means that the accretion layer accretes more material, the outburst intensity is higher, and the low-energy nuclear neutrino luminosity produced during the outburst is higher.

\begin{figure}
	\centering
		\includegraphics[width=\columnwidth]{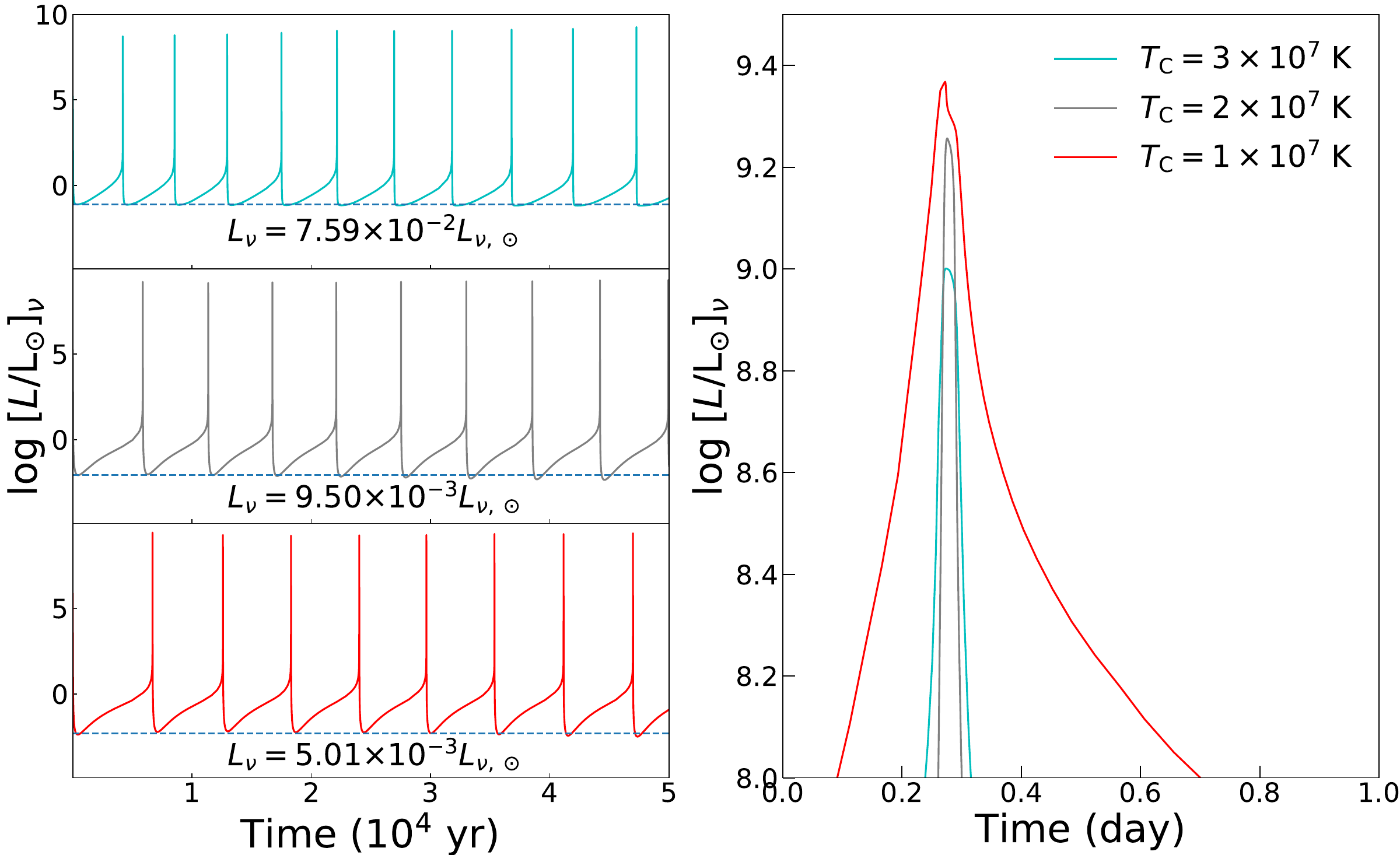}    
		\caption{Neutrino luminosity curves with different $T_{\mathrm{C}}$. Left:
		the neutrino luminosity of the burst sequence during 0$-$50000 yr with different $T_{\mathrm{C}}$, i.e., $1\times10^{7}$ (red), $2\times10^{7}$ (gray), $3\times10^{7}\rm \ K$ (cyan). Right: average light curves.
		Both nova models have accretion rate $1\times10^{-8} \ \mathrm{M}_{\odot}\rm\ yr^{-1}$, and accrete $25\%$ WD material and $75\%$ solar material at a 0.9 $\rm M_{\odot}$ CO WD.}
		\label{fig:6}
	\end{figure}

The left panel of Fig. \ref{fig:6} displays the neutrino luminosity curves of a hot model $(T_{\mathrm{C}}=3\times10^{7})$ and cold models $(T_{\mathrm{C}}=2\times10^{7}\ \mathrm{and}\ T_{\mathrm{C}}=1\times10^{7}\ \rm K)$ and with $0.9M_{\odot}$ WD. The right panel shows the average peak low-energy nuclear neutrino luminosity of ten outbursts.
When the initial core temperature of the nova model WD is lower, most of the gravitational energy released by the accretion compressed material will be directed to the core, which will delay the time to reach the TNR. Of course, it means the envelope will accrete more H-rich materials, the outbursts are more intense, and the low-energy nuclear neutrino luminosity is higher. However, due to the lower temperature of the core and accreted layer, the nuclear reaction rate during accretion is also lower, resulting in a low neutrino luminosity during the accretion phase.  
In the hot WD $(T_{\mathrm{C}}=3\times10^{7}\ \rm K)$, convection zones will appear in the outer core layer. The accreted hydrogen rapidly mixes in the convective region and continues to diffuse to the core, leading to an early outburst \citep{Schwartzman1994}. However, because the accretion mass is less and the temperature is lower, the low-energy nuclear neutrino luminosity produced by the outburst is lower.

\begin{figure}
	\centering
		\includegraphics[width=\columnwidth]{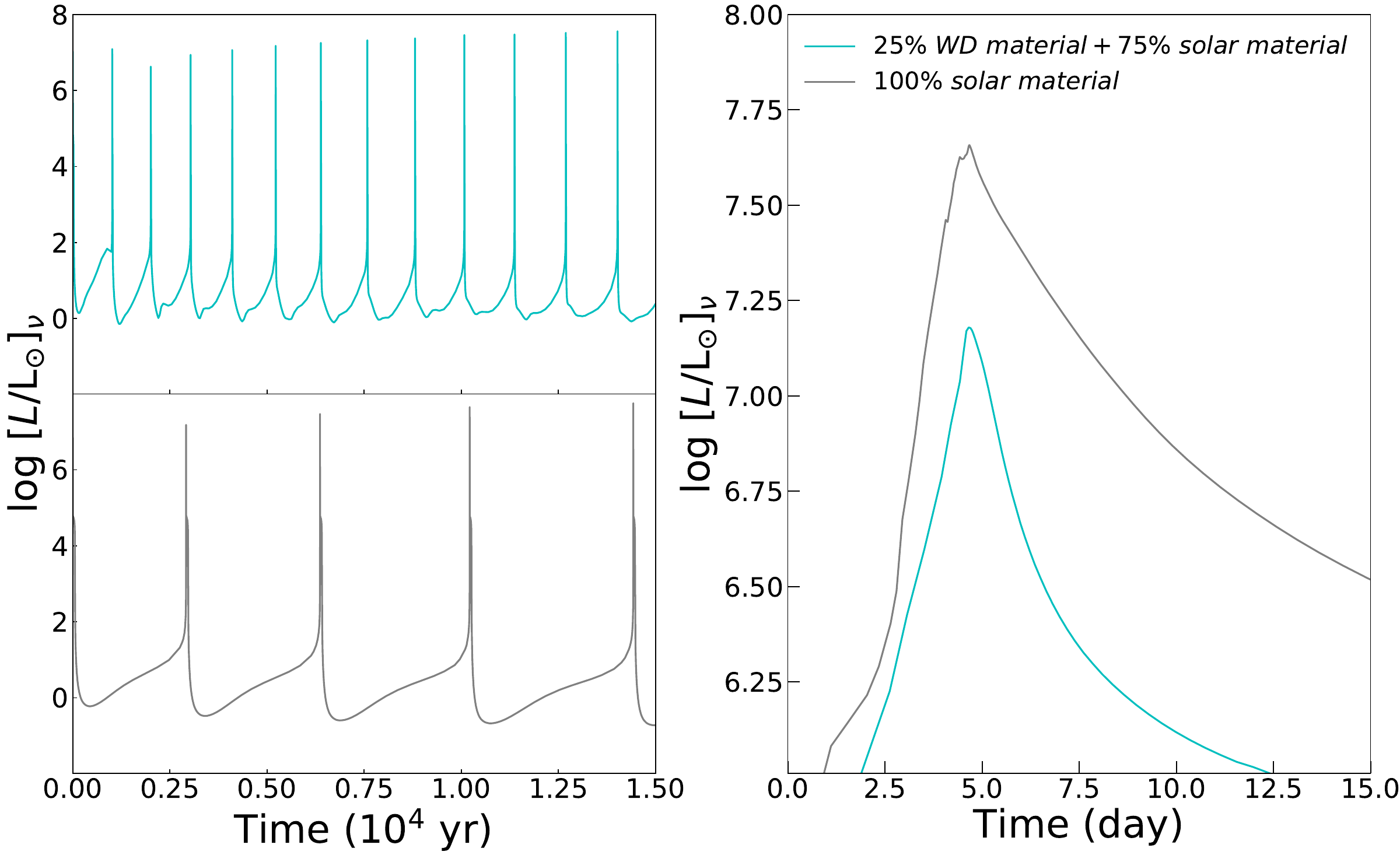}    
		\caption{Neutrino luminosity curves with different mixing. Left:
		the neutrino luminosity of the burst sequence during 0$-$15000 yr when accreting $25\%$ WD material and $75\%$ solar material and not considering any mixing. Right: average light curves.
		Both nova models have center temperature $T_{\mathrm{C}}=3\times10^{7}\ \rm K$, and 0.8 $\rm M_{\odot}$ CO WD at a rate $\dot{M}=1\times10^{-8} \ \mathrm{M}_{\odot}\rm\ yr^{-1}$.}
		\label{fig:7}
	\end{figure}

The left panel of Fig. \ref{fig:7} shows the neutrino luminosity curves of nova models when mixing is considered or not. The right panel shows the average peak low-energy nuclear neutrino luminosity of ten outbursts.
If the mixing of accretion material and core CO-enriched material is considered during the accretion stage, it means an increase in the metallicity of the accretion material. On the one hand, increasing the metallicity of the accreting material results in an increase in the opacity. The increased opacity not only allows more heat from the compression of accretion material to be trapped in the accretion layer, but also more heat from nuclear burning to be trapped in the nuclear reaction zone so that the temperature in the nuclear burning region increases faster and the time to reach TNR is significantly shorter than not considering the mixing nova model. Therefore, the total amount of accreted and ejected mass is less \citep{Starrfield1998}. On the other hand, mixing of CO-enriched materials can also outburst earlier because it acts as a catalyst for the CNO cycle (H burning). 
Models that do not take into account mixing require more accretion matter because the temperature rises more slowly in the nuclear burning region.
Thus, compared to nova models that consider mixing, the nova model that does not consider mixing has a denser accretion layer, more intense outbursts, and higher low-energy nuclear neutrino luminosity.

\subsection{Neutrino luminosity of T CrB nova}
\label{subsection:3.3}
\begin{figure*}
    \centering
        \includegraphics[width=\textwidth,height=0.4\textwidth]{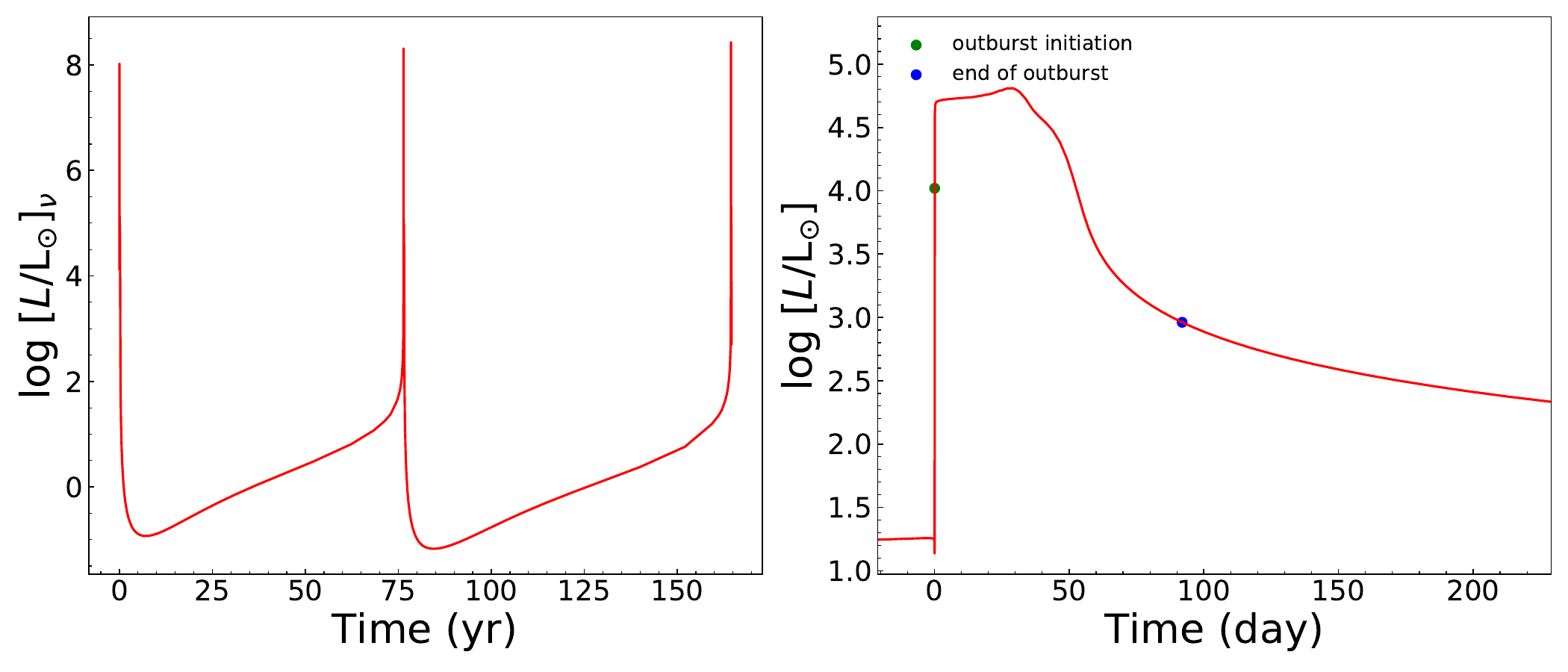}    
        \caption{The neutrino luminosity (left) and photon luminosity (right) curves of T CrB nova. Left: the first two bursts of the T CrB nova and the next one shortly after. Right: predicted duration of the next outburst of T CrB.}
        \label{fig:8}
    \end{figure*}

Considering $M_{\rm WD}$ of T CrB is 1.1$-$1.3$\ \rm M_{\odot}$ and the time span between outbursts is approximately 80 yr, the average accretion rate $\dot{M}_{a}\approx2\times10^{-8}\ \mathrm{M}_{\odot}\rm yr^{-1}$ \citep{Zamanov2023}, so we use $M_{\mathrm{WD}}=1.3\ \rm M_{\odot}$, $\dot{M}=1.72\times10^{-8}\ \mathrm{M}_{\odot}\rm\ yr^{-1}$ as our T CrB nova model. The left panel of Fig. \ref{fig:8} simulates the low-energy nuclear neutrino luminosity of the first two bursts of the T CrB and the next one shortly after, with peak low-energy nuclear neutrino luminosity of $1.06\times10^{8}$, $2.05\times10^{8}$, and $2.70\times10^{8}\ L_{\nu,\odot}$, respectively. In addition, according to the definition of outburst duration in Sec. \ref{sec:2}, it can be predicted that the next outburst of T CrB will last about 88 days, as shown in the right panel of Fig. \ref{fig:8}.

\subsection{Neutrino energy spectrum}
In this section, we consider the neutrino spectrum is produced in two ways by novae: one by nuclear reactions inside novae and the other by hadronic processes outside novae.

\subsubsection{Nuclear reaction}
Consider that the scene of the nova outburst is similar to that of supernovae explosion. The CNO cycle that dominates nova outbursts and the fast electron-neutrino bursts $(\mathrm{p}+\mathrm{e}^{-} \longrightarrow \mathrm{n}+v_{\mathrm{e}})$ in supernovae are both $\beta$ decay, so we think that the neutrino spectrum produced by the nova nuclear reaction is similar to the supernova neutrino spectrum. Following \citet{Baxter2022}, the initial neutrino energy spectrum has the form

\begin{equation}
	\Phi\left(E_\nu\right)=\frac{L_\nu}{4\pi D^{2}_{\mathrm{TE}} \left\langle E_\nu\right\rangle} \frac{(\alpha+1)^{\alpha+1}}{\left\langle E_\nu\right\rangle \Gamma(\alpha+1)}\left(\frac{E_\nu}{\left\langle E_\nu\right\rangle}\right)^\alpha \exp \left(-\frac{(\alpha+1) E_\nu}{\left\langle E_\nu\right\rangle}\right) ,
	\end{equation}
where $D^{2}_{\mathrm{TE}}$ is the distance between T CrB and Earth, and ${\left\langle E_\nu\right\rangle}$ is mean energy. The $\alpha$ is a spectral shape parameter.

\begin{figure}
    \centering
        \includegraphics[width=\columnwidth]{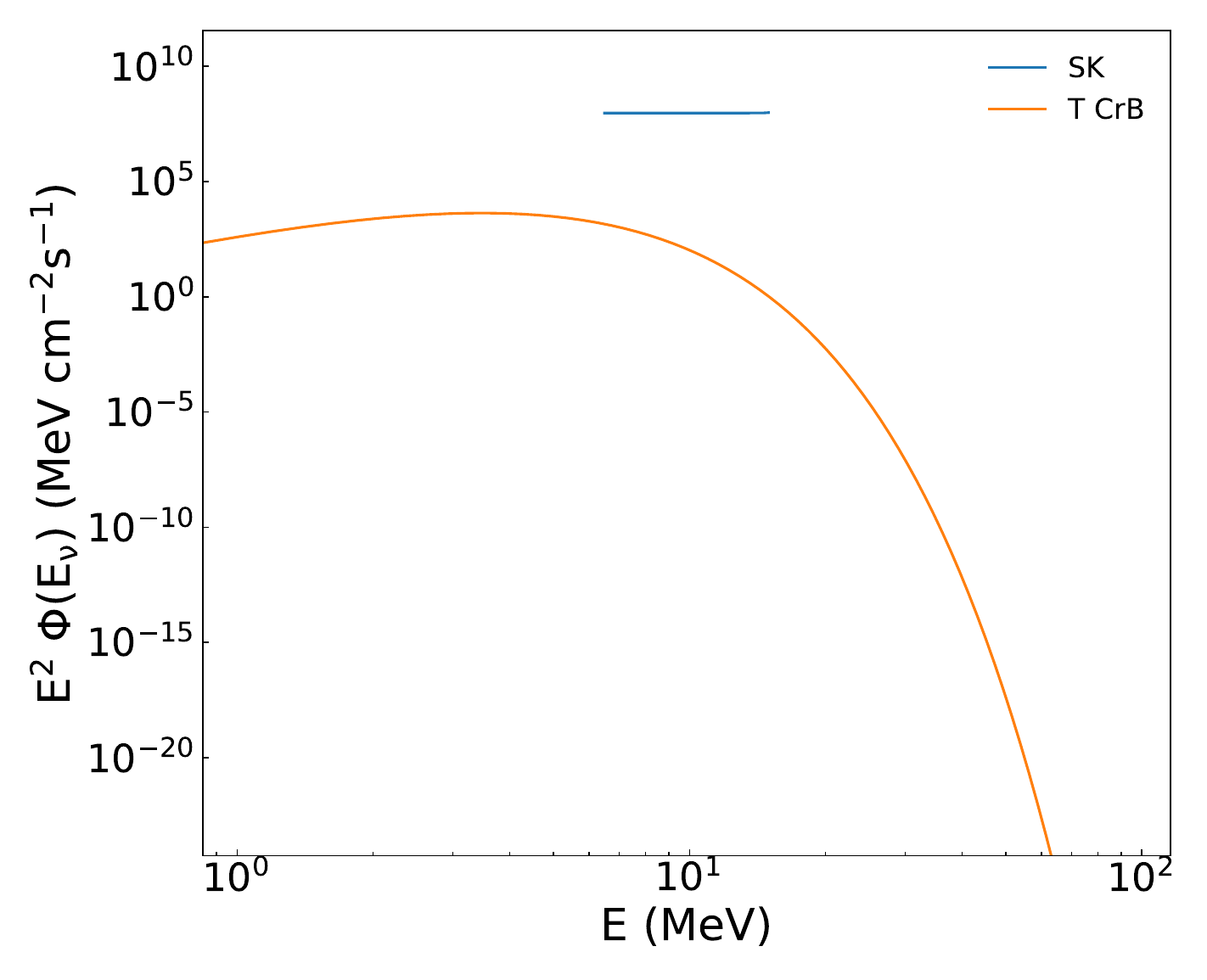}    
        \caption{The estimated total electron neutrino flux reaching Earth from T CrB (orange line). The Super-Kamiokande's upper limit on electron neutrino flux \citep{Suzuki2019} is the blue line.}
        \label{fig:9}
    \end{figure}

As discussed in Sec. \ref{subsection:3.3}, the low-energy nuclear neutrino luminosity of the next outburst of T CrB is $2.70\times10^{8}\ L_{\nu,\odot}$. The CNO cycle produces neutrinos with mean energies of about 2 MeV \citep{BorexinoCollaboration2020}. The $\alpha$ is 2.5 \citep{Keil2003}. The electron neutrino spectrum is shown in Fig. \ref{fig:9}. The model predicts the electron neutrino flux from nuclear reaction of T CrB outburst that cannot exceed the sensitivity limit of Super-Kamiokande. 

\subsubsection{Shock speed}
\label{subsection:3.4.2}
Protons accelerated by the shock of the nova interact with ambient particles in the RG wind can produce high-energy neutrinos. The flux of muon neutrinos reaching Earth from the T CrB is calculated using the semianalytical formulation developed in \citet{Kelner2006},

\begin{equation}\label{equ:18}
	\Phi_\nu\left(E_\nu\right)=\frac{\mathrm{c} n_{\mathrm{R G}}}{4\pi D^{2}_{\mathrm{TE}}} \int_0^1 \sigma_{\text {inel }}\left(E_\nu / x\right) J_p\left(E_\nu / x\right) F_\nu\left(x, E_\nu / x\right) \frac{d x}{x} ,
\end{equation}
where c is light speed.
$F_\nu\left(x, E_\nu / x\right)$ represents a neutrino spectrum produced by a single energy proton, and $x$ is $E_{\nu}/E_{p}$. $\sigma_{\mathrm{inel}}$ is total cross section of $pp$ interactions,
\begin{align}
	&&& \sigma_{\rm inel}(E_{\mathrm{p}})=34.3+1.88 L+0.25 L^{2}\rm \ mb\ ,
\end{align}
where $L=\mathrm{ln}(E_{\mathrm{p}}/1\ \mathrm{TeV})$. $J_{\mathrm{p}}\left(E_\nu / x\right)$ represents the distribution of protons

\begin{align}
	&&& J_{\mathrm{p}}\left(E_{\mathrm{p}}\right)=\frac{A}{E_p^\alpha} \exp \left[-\left(\frac{E_{\mathrm{p}}}{E_0}\right)^\beta\right]\ ,
\end{align}
where $\alpha$ is the power-law index, which we select to be 2.2, and $\beta=1$ \citep{Kelner2006}. $A$ is determined based on the given condition:

\begin{align}
	&&& \int_{E_{\mathrm{min}}}^{E_{\mathrm{max}}} E_{\mathrm{p}} J_{\mathrm{p}}\left(E_{\mathrm{p}}\right) d E_{\mathrm{p}}=E_{\rm {dens,T\ CrB }}\ ,
\end{align}
where the minimum proton energy is the rest mass energy of the proton ($E_{\mathrm{min}}\approx1\rm\ GeV$). The $E_{\rm {dens,T\ CrB }}$ is the energy density of protons around T CrB, determined by the kinetic energy of the shock wave \citep{Acciari2022},

\begin{equation}
    E_k=0.5 M_{e j} v_{s h}^2=2.0 \times 10^{44}\left(\frac{M_{e j}}{10^{-6} M_{\odot}}\right)\left(\frac{v_{s h}}{4500 \mathrm{~km} \mathrm{~s}^{-1}}\right)^2 \mathrm{erg}\ ,
\end{equation}
where $M_{\mathrm{ej}}$	is the total ejection mass during the eruption. Based on our T CrB nova model, $M_{\mathrm{ej}}=1.36\times10^{-6}\ \rm M_{\odot}$. According to \citet{Caprioli2014}, $10\%$ of shock kinetic energy can be converted to the energy of high-energy protons. For the T CrB nova, the energy released into accelerated protons ($E_{\rm p,nova}$) is about $2.72\times 10^{43}\ \rm erg$. Meanwhile, the energy density of protons surrounding the T CrB can be determined by dividing the total energy by the volume of the corresponding region,

\begin{equation}
    E_{\rm {dens,T\ CrB }}=\frac{3 E_{p, \text { nova }}}{4 \pi r_{\text {sh}}^3}\ .
    \end{equation}

Finally, we can calculate that the proton energy density near the shock wave is $102\ \rm erg/cm^{3}$, which is much greater than that of the average energy density in cosmic rays in the Milky Way ($\sim\ 1\ \rm erg/cm^{3}$).

\begin{figure}
    \centering
        \includegraphics[width=\columnwidth]{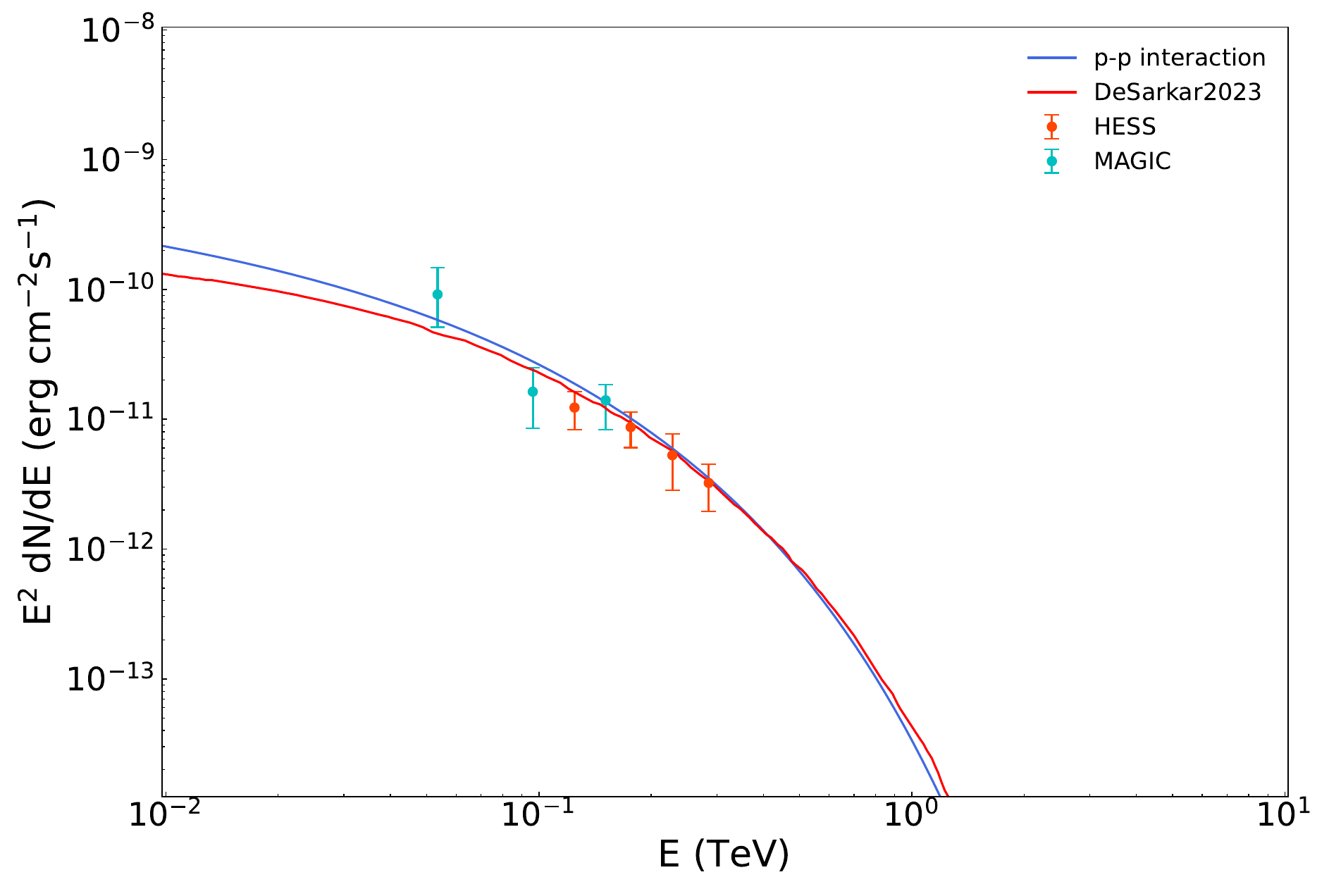}    
        \caption{The estimated $\gamma$-ray flux produced by the $pp$ interaction from the RS Oph outburst. The orange line is the $\gamma$-ray flux produced by the $pp$ interaction from the RS Oph outburst predicted by \citet{DeSarkar2023}. The blue line is the total muonic neutrino flux from the RS Oph outburst by our hadronic model. H.E.S.S. and MAGIC data points are taken from \citet{Aharonian2022} and \citet{Acciari2022}, respectively.}
        \label{fig:10}
    \end{figure}

To validate the credibility of our hadron model, we conduct an estimation on the $\gamma$ rays emitted during the RS Oph nova event and compare our results with those presented by \citet{DeSarkar2023}.
Protons accelerated by the shock of the nova interacting with ambient particles in RG wind also can produce high-energy $\gamma$-ray, $pp$ interactions resulting in the generation of neutral muons; these muons subsequently decay into high-energy $\gamma$-ray 
($\pi^{0}\to2\gamma$). The fitting formula of $\gamma$-ray flux produced by $pp$ interaction has the same form as Eq. \eqref{equ:18},

 \begin{equation}
	\Phi_\gamma\left(E_\gamma\right)=\frac{\mathrm{c} n_{\mathrm{R G}}}{4\pi D^{2}_{\rm RE}} \int_0^1 \sigma_{\text {inel }}\left(E_\gamma / x\right) J_p\left(E_\gamma / x\right) F_\gamma\left(x, E_\gamma / x\right) \frac{d x}{x}\ ,
\end{equation}
where $D_{\rm RE}$ is the distance between RS Oph and Earth, and $x$ is $E_{\gamma}/E_{p}$. The key parameters for calculating the $\gamma$-ray emissions of RS Oph include $v_{\mathrm{sh}} = \rm 4500\ km/s$, $r_{\mathrm{sh}} = \rm 4\times10^{13}\ cm$, $v_{\mathrm{RG}} = \rm 10\ km/s$, $\dot{M}_{\mathrm{R G}} = 5\times10^{-7}\ \rm M_{\odot}\ \mathrm{yr}^{-1}$, $\tau_{p p} \approx 2.2 \times 10^6 \ \mathrm{s}$, $B=0.11\ \rm G$, $\alpha = 2.2$, and $E_{0}=400 \ \rm GeV$. The $\gamma$-ray flux from RS Oph to Earth is depicted in Fig. \ref{fig:10}. Our analysis reveals that the MAGIC and H.E.S.S. $\gamma$-ray data points can be consistently explained by $\gamma$ rays generated within our hadronic model.

\begin{figure}
    \centering
        \includegraphics[width=\columnwidth]{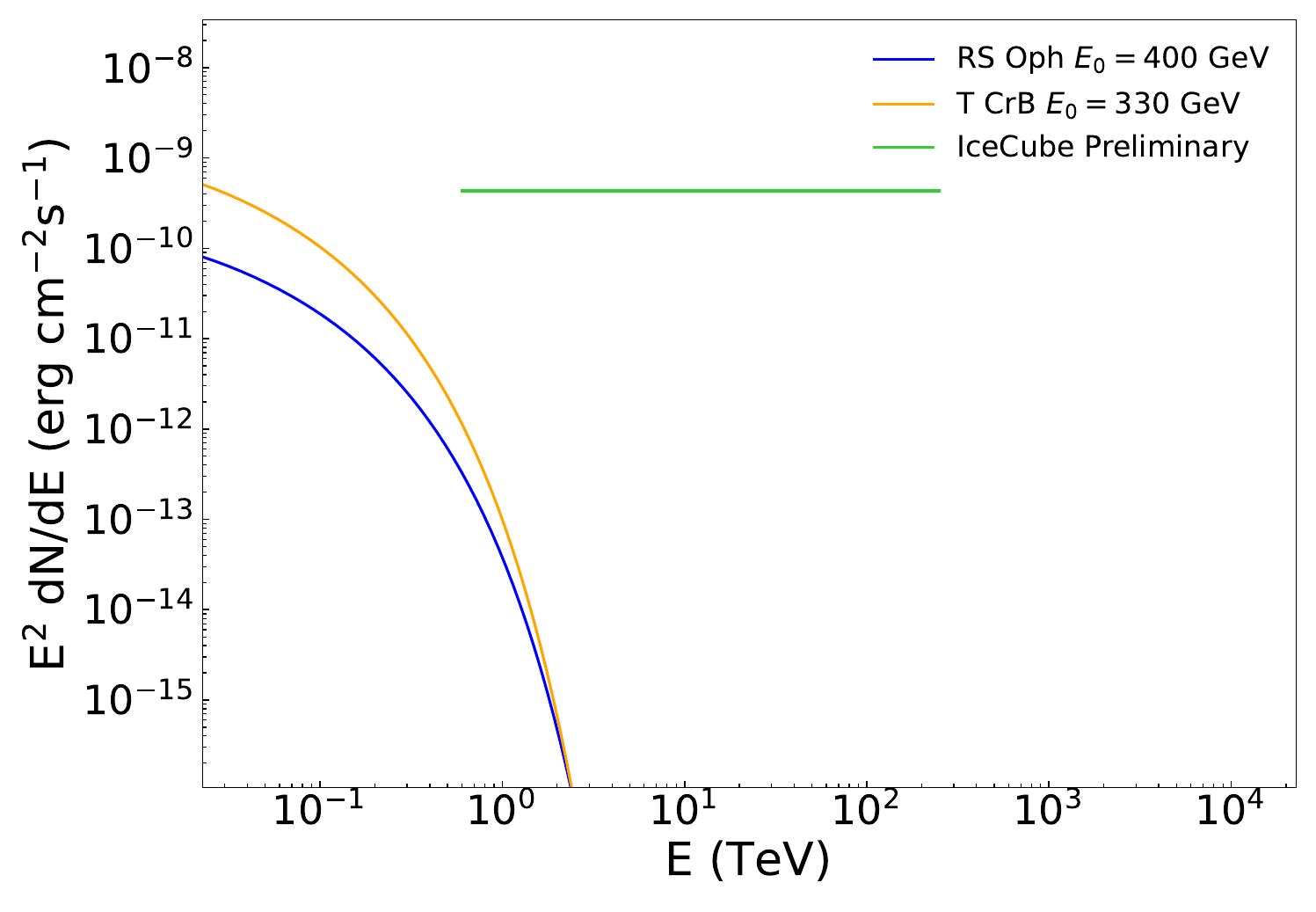}    
        \caption{The estimated total muonic neutrino flux reaching Earth from T CrB (orange line). The blue line is the total muonic neutrino flux from the RS Oph outburst. IceCube's upper limit on muon neutrino flux is the green line \citep{Pizzuto2021}.}
        \label{fig:11}
    \end{figure}

The muon neutrino flux after considering neutrino oscillations is illustrated in Fig. \ref{fig:11}. Neutrino oscillations refer to the phenomenon where neutrinos change from one flavor state to another during propagation, including transformations between electron neutrinos, muon neutrinos, and $\tau$ neutrinos. That is, the muon neutrino flux reaching Earth is 1/3 of the total neutrino flux produced by the $pp$ interactions at the T CrB. The blue line is the neutrino flux from the RS Oph outburst predicted by \citet{DeSarkar2023}. We reproduce their result and compare it with our predictions of the neutrino flux of the T CrB. The neutrino spectrum generated by the T CrB outburst exhibits similarities to that of RS Oph, as predicted by \citet{Gagliardini2024}. However, due to its closer proximity and higher maximum energy of proton production during the outburst, the resulting neutrino flux reaching Earth is amplified.
As shown in Fig. \ref{fig:11}, it is unlikely that the current-generation IceCube will observe it.

\section{Conclusions} \label{sec:4}
Using \textsc{mesa}, we simulate the evolution for nova outburst with CO WD mass ranging from 0.6 to 1.1 $\rm M_{\odot}$, and 1.1 $\rm M_{\odot}$ ONe WD. We calculate the low-energy nuclear and thermal neutrino luminosity and give neutrino H-R diagram as well as low-energy nuclear and thermal neutrino luminosity curves. Our results show that, 
during the accretion phase, the nuclear neutrino is mainly produced by the pp chains, and the thermal neutrino is mainly dominated by plasma decay. Neutrinos are produced mainly by thermal processes. Compared with photons, neutrinos carry away less energy.
However, during the TNR, the temperature of the nuclear reaction zone on the surface of the nova exceeds $7\times10^7\rm\ K$, the nuclear neutrino is mainly dominated by the CNO cycle, and the thermal neutrino is mainly dominated by photon neutrino, and the nuclear neutrino will far exceed the thermal neutrino. Meanwhile, the photon is easily trapped during TNR, and the low-energy nuclear and thermal neutrino luminosity rises rapidly and is much larger than the photon luminosity. 

The $M_{\mathrm{WD}}$, $\dot{M}$, $T_{\mathrm{C}}$, and mixing have significant effects on the low-energy nuclear neutrino luminosity of novae. 
The temperature of the nuclear reaction zone on the surface of a WD increases with mass, so does the low-energy nuclear neutrino luminosity. 
The smaller the $\dot{M}$, the longer the outburst interval, the more accreted material, and the more intense the outburst, so the greater the low-energy nuclear neutrino luminosity. 
The model that does not take into account mixing has a higher low-energy nuclear neutrino luminosity because it has a denser accretion layer and a higher temperature.  
The cold nova model has lower low-energy neutrino luminosity in the accretion phase and higher low-energy nuclear neutrino luminosity during the TNR. This is because its core and shell are both cooler during the accretion phase. It takes longer to reach the TNR, so more material is accreted, the shell is hotter, and the eruptions are more intense.

We predict that the low-energy nuclear neutrino peak luminosity of the next outburst of the T CrB will be $2.7\times10^{8}\ L_{\nu,\odot}$ with a low-energy nuclear neutrino outburst duration of 88 days. We also predict the low-energy nuclear and high-energy hadronic neutrino flux that will reach Earth during the next outburst of T CrB. The results show that it is unlikely for current-generation IceCube to observe the neutrino flux produced from the hadronic processes during the T CrB next outburst.

\begin{acknowledgments}
This work received the generous support of 
the National Natural Science Foundation of 
China under Grants No. 12163005, No. U2031204, No. 12373038, No. 12288102, and No. 12263006, the 
science research grants from the China Manned Space Project with No. CMSCSST-2021-A10, 
the Natural Science Foundation of Xinjiang No. 2022D01D85 and No. 2022TSYCLJ0006, 
and the Major Science and Technology Program of Xinjiang Uygur Autonomous Region under Grant No. 2022A03013-3.
\end{acknowledgments}

\section*{Data Availability}
The data that support the findings of this article are not publicly available. The data are available from the authors upon reasonable request.

\bibliography{sample631}{}

\end{document}